\documentclass[12pt]{article}

\usepackage{amsfonts,amsmath,amsxtra}
\newcommand{\beq}[1]{\begin{equation}\label{#1}}
\newcommand\eeq {\end{equation}}
\newcommand\bqa {\begin{eqnarray}}
\newcommand\eqa {\end{eqnarray}}
\newcommand\pr {\partial}

\newcommand{\bear}{\begin{array}}
\newcommand{\enar}{\end{array}}
\newcommand{\hf}{\frac{1}{2}}

\newcommand{\K}{\mathbb{K}}

\begin{document}

\def\t{\theta}
\def\T{\Theta}
\def\w{\omega}
\def\ov{\overline}
\def\a{\alpha}
\def\b{\beta}
\def\g{\gamma}
\def\s{\sigma}
\def\l{\lambda}
\def\wt{\widetilde}

%\begin{flushright}
% RG-ADS.tex
%\end{flushright}

\hfill ITEP-TH-05/02

\vspace{10mm}

\centerline{\Large \bf Notes on Multi-Trace Operators and}
\centerline{\Large \bf Holographic Renormalization Group\footnote{Talk
given
at "Particle, Strings and Fields" Vancouver, July 2000; "30 years
of SUSY" Minneapolis, October, 2000; "Integrable Models, Strings and
Quantum
Gravity", Cennai, January 2002. Based on the unpublished work
done in collaboration with A.Gerasimov.}.}

\vspace{5mm}

\centerline{\bf E.T. Akhmedov}

\centerline{117259, ul. B.Cheremushkinskaya, 25, ITEP, Moscow}

\vspace{5mm}

\begin{abstract}
It is shown that the Holographic Renormalization Group can
be formulated universally within Quantum Field Theory
as (the quantization of) the Hamiltonian flow on the
cotangent bundle to the space of gauge-invariant single-trace
operators supplied with the canonical symplectic structure.
The classical Hamiltonian dynamics is recovered
in the large $N$ limit.
\end{abstract}

\vspace{5mm}
%\section{Introduction}

{\bf 1.} There are many places in modern Quantum Field Theory
in which one finds the relation as follows:

\bqa \label{main}
\left\langle \exp\left\{{\rm i}\, \int d^D x \, g^n(x) \,
O_n\left[\Phi(x)\right] \right\}\right\rangle
\propto \Psi^{(D+1)}_{\cal N}\left[g(x)\right],
\eqa
where in the LHS the average is taken
with a weight $\exp\{{\rm i} \, S_0[\Phi]\}$ in a $D$-dimensional
boundary
Quantum Field Theory (bQFT). The content of fields in bQFT
is denoted for simplicity by $\Phi$,
$O_n[\Phi]$ is a basis of local gauge invariant
operators in the theory. At the same time on the RHS of (\ref{main})
$\Psi^{(D+1)}_{\cal N}$ is a wave function in a $(D+1)$-dimensional Bulk

Quantum Field Theory (BQFT) corresponding to a quantum number
${\cal N}$.

Known examples of such a relation are:

\begin{itemize}

\item The relation between three-dimensional Chern-Simons
theory and two-dimensional WZNW model \cite{WitCS};

\item AdS/CFT correspondence and its relatives \cite{Mal,GuKlPo,Wit};

\item As well some what similar is the relation
between two-dimensional conformal topological models
and old matrix models (see e.g. \cite{Morozov});

\end{itemize}

In these examples it appears that supplying the fields in bQFT
with $(N\times N)$ matrix indexes gives a natural parameter
in the $(D+1)$-dimensional BQFT. In fact, the phase space of the
BQFT theory is given by the functionals
on the space of the gauge-invariant operators $O_n$ and corresponding
sources $g^n$ of the theory. This space has a natural
symplectic structure. For instance in  the $D=0$ case there is the
duality between coupling constants $\{g^n\}$ and the traces of the
powers of the fundamental field $O_n=\{\frac{1}{n}Tr\Phi^n\}$. In
the limit $N\rightarrow \infty $ these are dual sets of variables
connected by Fourier transform (Legandre transform
for arbitrary $D$) defined by the integral kernel:

\bqa\label{legand}
Z\left(\{g\}^{\phantom{\frac12}};\{O(\Phi)\}\right)
= \sum e^{\sum_{n=0}^{\infty} g^n O_n}
\eqa
Note that if we do not take the large $N$ limit there
is a finite number of independent $O_n$ and infinite
set of $g^n$.

{\bf Hence, the configuration space $\{g^n\}$ is a linear vector
space and the symplectic structure

\bqa \label{simpl}
\omega = \delta g^n \wedge \delta O_n
\eqa
is non-degenerate only in the limit $N \to \infty$}.
Note that as well $g^n$ and $\Phi$ are good coordinates
on the phase space, but in the $g^n$ and $O_n$ coordinates
the symplectic structure takes the simplest form (\ref{simpl}).

More specifically, there are three regimes.
First regime is when $N$ is finite,
then we can always take finite number of $g^n$ and the
symplectic structure (\ref{simpl}) is non-degenerate. However, it seems
that
this corresponds to the situation when the configuration space
is not a linear space. Hence, the aforementioned kernel
of the Fourier transform is not defined in the simple
form (\ref{legand}). Second regime is when
$N = \infty$ and the configuration space becomes
a linear vector space with the well defined Fourier kernel
of the type (\ref{legand}). Moreover, in this case
we obtain the classical approximation for BQFT \cite{Mal}.
Third regime is when we expand around $N=\infty$, i.e. $N\to\infty$ but
finite. Then we obtain the quantization of the second regime.

What is most important for us is that the formula (\ref{main}) reveals
the
deep relation of the {\it Euclidian}
time evolution in BQFT and the conformal properties of the
bQFT \cite{Mal,Polyakov}.
In other words there is a relation between equations
of motion in BQFT and Renormalization Group (RG) flow in bQFT
\cite{Akh,Verlinde,HRG}.

In particular it is tempting
to connect the effective action at the
momentum scale $p^2= M^2(u)$ in bQFT and the wave function at
the constant "time" slice $u=const$ in BQFT. The function $M(u)$ is
given by the solution of the classical equations of motions of
BQFT. When this function is invertible (as in the AdS/CFT case) it
is natural to identify the scale of the boundary theory with the
"time" coordinate in BQFT. In the AdS/CFT
correspondence the connection is \cite{Mal}:

\bqa
\label{u}
e^{\phi (u)}\propto \frac{u}{R}.
\eqa
(here $R$ is the "radius" of the AdS space --- UV regulator for
D-dimensional gauge theory. This connection may be obtained
from the requirement of conformal invariance
of the term $\sqrt{det(h)}Tr[ \Phi,\Phi]^2$ in
the action of $D=4$ $N=4$ SYM.) However  in general the
transformation is more involved. The simplest
non-trivial case is the SUGRA solution corresponding to several
groups of D-branes \cite{Mal,KlWi}.

This conjectured connection between quantum theories in different
dimension gives rise to the following natural questions:

\begin{itemize}

\item {\it What is the correct configuration space $\{g^n\}$, i.e.
generating function of what kind of operators in bQFT is
appropriate for the relation (\ref{main})?}

\item {\it How RG equations in bQFT, which are normally first order
differential equations, become second order differential classical
equations in BQFT.}

\item {\it How RG flow becomes reversible? In fact, one have
to relate somehow seeming irreversibility of RG flows with
the Holography.}

\item {\it What is the meaning of the quantum
number ${\cal N}$ from the point of view of bQFT?
Or more generally, what is the meaning of an arbitrary
wave-function in BQFT in terms of bQFT?}

\end{itemize}

The purpose of this note is to give answers to these
questions via consideration of a concrete example of bQFT.

{\bf 2.} We begin with some simple remarks on the
RG flow in field theory. For definiteness from now on
we consider the matrix field theory

\bqa \label{lagr}
S_0 = \int d^D x \, N \, Tr \left(|\pr_{\mu} \Phi|^2
+ \frac12 m^2 |\Phi|^2 + g^{(4)} |\Phi|^4\right),
\eqa
where $\Phi$ is a field taking values in the
adjoint representation of $SU(N)$.

Consider RG flow in this theory in the
Wilson's picture \cite{WiKo}. The basic ingredient of the Wilson's
approach is the effective action defined at some scale.
To obtain the latter,
one separates the field $\Phi$ into slow classical
modes $\Phi_0$ dependent on the momenta $p^2 < u^2$ and
quantum fluctuations $\varphi$ dependent on the momenta $p^2 > u^2$:
$\Phi = \Phi_0 + \varphi$.
Then one takes the functional integral over $\varphi$.
Along this way one obtains the effective action which
is a linear combination of all possible operators in the theory
\cite{Pol}:

\bqa
\label{action}
S(g^n,O_n) = \iint d^D x d^D y \, N \, K(x,y\mid u) \,
Tr \left[\Phi(x)^{\phantom{\frac12}} \Phi(y)\right] +
\int d^Dx \, g^n \, O_n[\Phi]
\eqa
where $K(x,y\mid u)$ is a regularized propagator in the
theory (\ref{lagr}). Note that the index $n$ in this formula
can include $D$-dimensional tensor indexes as well.

Thus, the field theory at the
normalization point $p^2\sim u^2$ is characterized by the coupling
constants dependent on the momenta $p^2>u^2$ and
the "classical" fields dependent on
the harmonics with the momenta $p^2<u^2$.
One can change the normalization point by
integrating out some of the classical field modes $\Phi_0$.
{\bf Note that by keeping the whole infinite set of $g^n$'s,
one keeps information about all higher frequency modes
$\varphi$, i.e. makes the RG flow reversible.}
Actually, as we show below it is enough to keep
only some infinite subset of all possible $g^n$'s.

Having in mind that $g^n$ and $O_n$ are coordinates on a phase space
it is natural to establish the relation of the type (\ref{main}).
In fact, the exponent of the action in the theory with the given bare
sources and given background fields averaged over the
fluctuations with the momenta $u_0^2>p^2>u^2$ may be interpreted
as the transition amplitude in a BQFT with "time-direction" along
$u$:

\bqa
\label{WIL}
\K\left(\{g^n(u_0)\}^{\phantom{\frac12}}, \, \{O_n(\Phi_0,\,u)\}\right)
\equiv \left\langle \{g^n\}\left| e^{\int^{u_0}_{u} du
\mathcal{H}(u)}\right|
 \{O_n(\Phi_0)\} \right\rangle =
\nonumber \\ = \int \prod_{u^2 < p^2 < u_0^2} d\varphi_p \,
\exp\left\{{\rm i} \, S_0[\Phi_0 + \varphi] + {\rm i} \,
\int g^n(x)\, O_n[\Phi_0 + \varphi]\right\},
\eqa
(in the AdS/CFT case $u_0 \propto R/\alpha'$) where $\mathcal{H}(u)$
is a Hamiltonian in a $(D+1)$-dimensional theory with dynamical
variables $g^n$ and $O_n$.
As we see, the wave function $\Psi^{(D+1)}$ can be expanded in the
basis of wave functions:

\bqa
\langle g^n| O_n\rangle = \exp\left\{{\rm i} \,
\int d^D x \, g^n \, O_n\right\},
\eqa
which appear from $\K$ when $u\to u_0$.

  At this point it is easy to see the meaning of the quantum number
${\cal N}$ in (\ref{main}). {\bf In fact, taking $u \to 0$, we observe
that $\Psi^{(D+1)}$ is characterized by quantum numbers ${\cal N} =
\{\langle O_n \rangle\}$, i.e. by the generalized momenta in the BQFT
or the VEV's of $O_n$'s in the bQFT.}
It is an interesting problem to find within bQFT
an explicit form of the operator which can
change ${\cal N}$ (inside the Hilbert space of BQFT).

   Most transparently  RG flow in the Wilson approach may be
represented in the form of the
Callan-Simanzik-Polchinski equations \cite{Pol}.
Now we are going to show that such equations can be represented as
Hamiltonian equations in a theory with the phase space
$\{g^n\}, \quad \{O_n(\Phi_0)\}$.

For the theory in question the Polchinski equation is as follows
\cite{Pol}:

\bqa\label{poleq}
\int d^D x \, \left(u\pr_u g^{n^{\phantom{\frac12}}}\right)\,
\left\langle O_n\left[\Phi_0 +
\varphi\right]^{\phantom{\frac12}}\right\rangle + \nonumber \\ +
\frac{1}{N} \,
\iint d^Dx d^Dy \, \left( u\pr_u K^{-1^{\phantom{\frac12}}}(x,y\mid u)
\right) \, \left\langle
e^{- {\rm i}\, S_1} \frac{\pr^2}{\pr \Phi^{ij}(x)\pr \Phi^{ji}(y)}
e^{{\rm i}\,S_1}
\right\rangle = 0
\eqa
where $S_1 = \int g^n \, O_n(\Phi_0 + \varphi)$ and the quantum average
is taken over $\varphi$ with the weight $\exp\{{\rm i}\, S_0\}$.

Using decomposition of the gauge invariant operators,
we obtain (similarly to \cite{Li}):

\bqa
\label{beta}
u \pr_u g^n \, O_n\left(\Phi_0\right) & - &\beta^n(g)
\, O_n\left(\Phi_0\right) - \nonumber \\ &-& \frac{1}{2N} \gamma^{nm}(g)
\,
O_n\left(\Phi_0\right) \, O_m\left(\Phi_0\right) + ...= 0,
\eqa
Here $\beta_n(g)$ and $\gamma_{nm}(g)$ are some {\it non-zero} model
dependent functions. It is straightforward to see that
$\beta$'s in (\ref{beta}) are $\beta$-functions for the
corresponding coupling constants.

There are two options now. If we use the full linear basis of the
operators in (\ref{action}) then, taking into account operator
product expansion:

\bqa
 O_n(x) \, O_m(y)|_{x\to y} = \sum_k C_{nm}^k \, O_k(x),
\eqa
we have a linear differential equation for the sources:

\bqa
u\pr_u g^n - \beta^n(g) - \gamma^{km}(g) \, C_{km}^n=0.
\eqa
This equation defines the RG flow on the sources in the theory.

However, if we use the basis in the space of the {\it single
trace} operators in
(\ref{action}), we obtain the differential
equations in the Hamiltonian form:

\bqa \label{hameq}
u \pr_u g^n & = & \beta^n(g)
+ \frac{1}{2N} \gamma^{nm}(g) \, Tr O_m\left(\Phi_0\right) \nonumber \\
u\pr_u Tr O_n\left(\Phi_0\right) & = &- \pr_n\beta^m(g) \, Tr
O_m(\Phi_0)
- \frac{1}{2N} \, \pr_n\gamma^{mk}(g) \, Tr O_m(\Phi_0) \,
Tr O_k(\Phi_0)
\eqa
where the second equation appears from (\ref{poleq})
when we vary it with respect to $g^n$. Note that we are explicitly
showing powers of traces in these equations.

Corresponding Hamiltonian function has the following form:

\bqa \label{Ham}
\mathcal{H}(g^n,\pi_m) = \beta^n(g) \, \pi_n + \frac{1}{2N}
\gamma^{nm}(g)\, \pi_n \, \pi_m
\eqa
and generates the RG-flow with respect to the
canonical symplectic structure (\ref{simpl})
($\pi_n \propto Tr O_n(\Phi_0) \propto \frac{\delta}{\delta\, g^n}$).
{\bf Thus, the proper choice of the configuration space $g^n$
is given by the sources for only single trace operators.}
Furthermore, as we noticed in the introduction,
the presence of the well defined symplectic structure
(\ref{simpl}) and well defined Legandre transform implies
the linearity (proper choice) of the configuration space.

Now the Hamiltonian (\ref{Ham}) defines the equation on the wave
function:

\bqa
\label{first}
\left[u\pr_u - \beta^n(g)\frac{\delta}{\delta g_n} - \frac{1}{2N}
\gamma^{nm}(g)\frac{\delta}{\delta g_n}
\frac{\delta}{\delta g_m} \right] \, \Psi(g,u)=0
\eqa
This equation is the right substitution of the RG-flow equations
for the nonzero $\gamma$. In the special case of vanishing $\gamma$
coefficients, we have:

\bqa
\left[u\pr_u - \beta^n(g)\frac{\delta}{\delta g^n}\right] \,
\Psi(g,u)=0,
\quad \Psi(g,u) = Z(g,u)
\eqa
Then the explicit solution of this equation is given
in terms of the RG-flow equations:

\bqa
 Z(g^n,u)&=&Z[g_*^n(g_n,u)]\nonumber \\
 u\pr_ug_*^n&=&\beta^n(g_*) \nonumber \\
 g_*^n(u\to \infty)&=&g^n
\eqa
Thus, in this special case we have the direct connection with
the standard RG flow.

{\bf It is interesting that combining scale transformation properties
of the theory with $N\rightarrow \infty$ limit we obtain
the classical equations of motion for the Hamiltonian defined
above instead of the first order differential equations.
Moreover these Hamiltonian equations are of higher order which is a
manifestation of the  appearance of multi-trace
operators.}

In the AdS/CFT correspondence the BQFT contains graviton as a
dynamical variable and thus formally has zero
Hamiltonian\footnote{Note that conformal invariance, which mixes
$x$'s and $u$ should restore general covariance in $D+1$-dimensions,
which is explicitly broken in the Hamiltonian approach.}. This
gives the Hamiltonian constraint on the wave function instead of
the evolution equation. This Hamiltonian constraint includes the
derivatives over the metric on a slice and thus contains the
information on the scaling properties of the bQFT. However the
Hamiltonian is non-linear over the momentum and thus gives rise to
the higher order equation on the metric. To reconcile this with
the equations discussed above one could use the energy-momentum
pseudo-tensor which may be defined in some fixed  coordinate
system. This new Hamiltonian gives rise to the "time" evolution
and was implicitly used in the calculations of
\cite{GuKlPo} (see the next section).

{\bf 3.}
Let us show here that AdS/CFT correspondence feats well
in to our picture.

Note first, that superconfomal theory of the SYM
constrains the corresponding BQFT uniquely (at least in the
case of 8 SUSY). This can not
be done so easily for other examples of bQFT
(note, at least, that the Hamiltonian
(\ref{Ham}) is defined perturbatively in $g^n$ through $\beta$'s
and $\gamma$'s.).

 AdS/CFT correspondence predicts the following
RG flow equation for the dilaton field in the linear approximation:

\bqa\label{solutions}
\left[z^3\pr_z \frac{1}{z^3} \pr_z +\Delta ^{(D)}\right] \, g = 0,
\eqa
where $\Delta^{(D)}$ is Laplace operator in $D$-dimensions and we
denote the dilaton field by $g$ to show the relation with our previous
considerations.
Corresponding action functional in the first order
formalism is given by:

\bqa \label{acjac}
S \propto \int \frac{dz}{z} \, \left[\pi \, z\pr_z g -\hf \, \pi^2 \,
z^4-
\frac{1}{2z^2} \, \pr_\mu g \, \pr^\mu g\right]
\eqa
Note that the corresponding evolution equations generate
quadratic in momenta
terms ($z\propto 1/u$, i.e. the UV limit $u\to \infty$
corresponds to $z\to 0$).

Equations  of motion in the Hamiltonian form are:

\bqa
  z\pr_z g=\pi z^4  \\
  \pr_z \pi =\frac{1}{z^3} \Delta^{(D)} g
\eqa
In the special case of $\Delta^{(D)}g=0$ we have:

\bqa
  z\pr_z g &=&\pi z^4  \\
  \pr_z \pi &=&0
\eqa
with the solution:

\bqa
 g&=&g_0+\pi_0z^4 \\
 \pi&=&\pi_0
\eqa
 ($g_0,\pi_0$ are integration constants).
More generally \cite{KlWi}:

\bqa
 g^{(\Delta)}\propto z^{D-\Delta}g^{(\Delta)}_0
+ z^{\Delta}\frac{1}{2\Delta-D} \pi^{(\Delta)}_0
\eqa
 Here we may note the mixing of the dual variables $g^n$ and
 $O_n$\,($\pi^n_0\propto O^n$).

   As we noticed in the previous section the action (\ref{acjac})
solves the Jacoby equation (see e.g. \cite{Verlinde}):

\bqa\label{constr}
\left[\frac{1}{\sqrt{G}}\left(\pi^{\mu\nu} \pi_{\mu\nu} - \frac13\,
\pi^\mu_\mu \, \pi^\nu_\nu\right) +
\frac12 \, \sqrt{G} \,\pi^2 + \frac12 \, \sqrt{G} \, \pr_\mu g \,
\pr^\mu g + \sqrt{G} \, {\cal R} \right] e^{- S_{min}} = 0,
\nonumber \\ \Psi_{cl}(g) = e^{- S_{\min}(g)}
\eqa
if minimized on the solutions of (\ref{solutions}). Here ${\cal R}$ is
four-dimensional curvature and $\pi^{\mu\nu}$
is the one conjugate to the component $G_{\mu\nu}$ of the metric:

\bqa
ds^2 = \left(d^{\phantom{\frac12}}\ln(z)\right)^2 +
G_{\mu\nu}(z,x) \, dx^\mu \, dx^\nu
\eqa
and we have set the AdS radius $R=1$ for simplicity.
In our case $G_{\mu\nu} = 1/z^2 \, \delta_{\mu\nu}$
and variations with respect to the metric are translated into
variations with respect to $z \propto \frac{1}{u}$.
Thus, along this way we obtain "time" evolution with respect
to $u$ from the Hamiltonian constraint (\ref{constr}) in the
theory with general covariance. Note that the equation
(\ref{constr}) is of the second order in $\pr_u$,
while (\ref{first}) is of the first order. The possible
resolution of this discrepancy is related to the restoration
of general covariance in the BQFT, which is explicitly broken
in the Hamiltonian formalism. Apriory, however, for arbitrary
bQFT the general covariance (in the corresponding BQFT) should
not be restored. For eaxmple, it could that the general covariance
is just explicitely broken by some VEV's
of tensor fields in BQFT.

{\bf 4.} It is worth mentioning that the described picture
is an attempt to combine the approach of loop equations \cite{Migdal}
and RG approach. Which was successfully done in the old matrix
models \cite{Morozov}. Apart from that a related to our considerations
is the appearance of the group theory interpretation of the RG flow
\cite{CoKr}.
Note as well the recent interest in the multi-trace
operators \cite{mltitr} and their relevance for the AdS/CFT
correspondence.

I would like to acknowledge discussions with I.Kogan, P.M.Ho, T.Wiseman,
H.Verlinde, A.Losev, A.Rosly, A.Gorsky, A.Morozov, A.Marshakov
and A.Young. Especially I would like to thank A.Gerasimov
for intensive collaboration.
This work was partially supported by the grant RFBR 01-01-00548 and by
the grant for support of young scientists 02-01-06360 and by
INTAS-00-390.


\begin{thebibliography}{50}

\bibitem{WitCS} E.Witten, Nucl. Phys. {\bf B311} (1988) 46.

\bibitem{Mal} J.Maldacena Adv. Theor. Math. Phys., {\bf 2} (1998) 231.

\bibitem{GuKlPo} S.Gubser, I.Klebanov and A.Polyakov, Phys. Lett.,
{\bf B428} (1998) 105.

\bibitem{Wit} E.Witten, Adv. Theor. Math. Phys., {\bf 2} (1998) 253.

\bibitem{Morozov} A.Morozov, hep-th/9502091; Phys.Usp., {\bf 37}
(1994) 1.

\bibitem{Polyakov} A.Polyakov, Nucl. Phys. Proc. Suppl, {\bf 68} (1998)
1.

\bibitem{Akh} E.Akhmedov, Phys. Lett., {\bf B442} (1998) 152;
hep-th/9806217.

\bibitem{Verlinde} J.de Boer, E.Verlinde and H.Verlinde, JHEP {\bf 0008}

(2000) 003; E.Verlinde, Class. Quant. Grav. {\bf 17} (2000) 1277.

\bibitem{HRG} E.Alvarez and C.Gomez, Nucl. Phys., {\bf B541} (1999) 441;

A.Morozov and A.Mironov, Phys. Lett. {\bf B490} (2000) 173.

\bibitem{KlWi} I.Klebanov and E.Witten, Nucl. Phys. {\bf B556} (1999)
89.

\bibitem{WiKo} K.Wilson and J.Kogut, Phys. Rept. {\bf 12} (1974) 75.

\bibitem{Pol} J.Polchinski, Nucl. Phys., {\bf B231} (1984) 269.

\bibitem{Li} M. Li. Nucl. Phys. {\bf B579} (2000) 525.

\bibitem{Migdal} A.Migdal and Yu.Makeenko, Nucl. Phys. {\bf B188} (1981)

269; Phys. Lett. {\bf B88} (1979) 135.

\bibitem{CoKr} A.Connes and D.Kreimer, Comm.Math. Phys.
{\bf 216} (2001) 215, Comm. Math Phys. {\bf 210} (2000) 249;
A.Gerasimov, A.Morozov and K.Selivanov, Int. J.Mod. Phys. {\bf A16}
(2001) 1531; D.Malyshev, hep-th/ 0112146.

\bibitem{mltitr} O.Aharony, M.Berkooz and E.Silverstein,
JHEP {\bf 0108} (2001) 006; E.Witten, hep-th/0112258.

\end{thebibliography}
\end{document}